# Physical Realization of a Hyper Unclonable Function

Sara Nocentini*[1,2], Ulrich Rührmair[3,4], Mauro Barni[5], Diederik S. Wiersma[1,2,6], Francesco Riboli*[2,7]

[1] Istituto Nazionale di Ricerca Metrologica, Strada delle Cacce 91, 10135 Torino, Italy; [2] European Laboratory for Nonlinear Spectroscopy, Via Nello Carrara 1, 50019 Sesto Fiorentino (FI), Italy; [3] Physics Dept. LMU Munchen, Schellingstraße 4/III D-80799 Munchen, Germany; [4] Electrical and Computer Engineering (ECE) Dept., University of Connecticut, Storrs, CT, USA; [5] Dipartimento di Ingegneria dell'Informazione e Scienze Matematiche, Università di Siena, via Roma 56, 53100 Siena; [6] Dipartimento di Fisica, Università di Firenze, Via Sansone 1, 50019, Sesto Fiorentino, Italia; [7] CNR-INO, Via N. Carrara 1 Sesto Fiorentino, 50019, Italy.

*nocentini@lens.unifi.it; riboli@lens.unifi.it

**Disordered photonic structures are promising materials for the realization of physical unclonable functions (PUF) – physical objects that can overcome the limitations of conventional digital security methods[1–6] and that enable cryptographic protocols immune against attacks by future quantum computers [7,8]. One PUF limitation, so far, has been that their physical configuration is either fixed or can only be permanently modified[9], and hence allowing only one token per device. We show that it is possible to overcome this limitation by creating a reconfigurable structure made by light-transformable polymers, in which the physical structure of the unclonable function itself can be reversibly reconfigured. We term this novel concept *Hyper PUF* or *HPUF* in that it allows a large number of physical unclonable functions to co-exist simultaneously within one and the same device. The physical transformation of the structure is done *all-optically* in a reversible and spatially controlled fashion. Our novel technology provides a massive enhancement in security generating more complex keys containing a larger amount of information. At the same time, it allows for new applications, for example serving multiple clients on a single encryption device and the practical implementation of quantum secure authentication of data[10].**

Complex photonic systems[11–17] are characterized by a multitude of spatial degrees of freedom that in presence of coherent light illumination produce in the far field a complex intensity pattern (called speckle pattern) as the result of the interference of a large number of independent transmission channels[18]. In particular, the optical speckle pattern that is generated by disordered materials is extremely sensitive to minute changes in the physical structure of the material[19,20], to the level that it is nearly impossible to clone such disordered structures and obtain the same optical response without resorting to cloning techniques at the molecular level. Such structural characteristics make them ideal candidates for cryptographic primitives such as physical unclonable functions for authentication and communication purposes [2,3]. Among the other types of PUFs, electrical Strong PUFs have been examined intensively by the PUF community [4,21,22], but most of them have been attacked successfully via various digital and physical techniques over the years[23,24]. Due to their promise of higher three-dimensional complexities and entropy levels, this has put optical PUFs back in the focus of recent PUF research.

Optical physical unclonable functions have been introduced by Pappu[6] with the name of Physical One-Way Functions. In this first instantiation, the PUF interrogation and the resulting challenge-response pair (CRP) protocol[4,13–15] relied on different angles of incidence of the laser and allowed to extract cryptographic keys with 230 independent bits (over a total bit string length of 2400 bits). While the optical setup based on moveable mechanical components limits the reproducibility of measurements, in later works the employ of modulators as challenge generators in the spatial[25,26] or spectral[27,28] domain provided a significant improvement. However, those PUFs rely on a static hardware whose properties cannot be reconfigured in case of detected attack. To overcome this limitation, Kursawe et al. showed that permanent modifications can be created by melting the polymer aggregates with a net entropy decrease in every new reconfiguration [29]. Horstmeyer and coauthors showed that it is possible to reconfigure an optical PUF by exploiting electrical driven polymer dispersed liquid crystals [25,23] and John et al. managed to do this electrically by using halide perovskite memristors[30]. In all these cases, the internal states of the PUF cannot be recovered after reconfiguration, and their entropy remains constant[25]. To increase the information entropy of the PUF, it is necessary to provide a reversible transformation among the possible microscopic configurations. A preliminary result in this direction was obtained by Gan and coauthors, who reported that the temperature-controlled phase transition of Vanadium oxides nanocrystals can be used to create a reversible switching among two states (crystalline and amorphous) [31].

In this context, we introduce a new concept and technology platform that provides interchangeable multi-level operation by reversibly transforming the scattering properties of a complex photonic medium based on photosensitive polymeric film. The operation principles of this cryptographic primitive – that we term Hyper PUF (HPUF) – is illustrated in Fig. 1a. A "standard" PUF (left panel of Fig. 1) is characterized by an authentication process via a single challenge $C_i^{Probe}$, while the HPUF (right panel of Fig. 1) is interrogated by a challenge $C_{ik} = (C_i^{Probe}, L_k^{Trans})$ consisting of two sub-challenges. First, a configuration pattern $L_k^{Trans}$ (a spatially modulated parametric matrix) transforms the internal configuration of the PUF between different levels in an all-optical and reversible manner. The configuration pattern determines the scattering potential. Each scattering potential is associated to a different level of the HPUF. Secondly, a standard interrogation challenge $C_i^{Probe}$ produces a measurable unique optical interference pattern as the PUF response $R_{ik}(C_i^{Probe}, L_k^{Trans})$. Mathematically, the HPUF can be modelled as a parametric function that maps its domain to a larger codomain, whose dimension depends not only on the number of $C_i^{Probe}$ but also on the number of transformer challenges $L_k^{Trans}$, i.e. $f: (C_i^{Probe}, L_k^{Trans}) \rightarrow R_{ik}$. The same internal configuration can be restored by applying the same transformer challenge, allowing back-and-forth switching between the PUF's internal levels. This marks a significant difference between HPUFs and existing reconfigurable PUF designs,[9,30,32] in which internal changes are permanent and non-reversible.

The practical usage of physical unclonable functions is governed by a registration and verification protocol of the challenge-response-pairs[33] that for standard and HPUFs differ in the library dimensionality and the type of challenge sent to the claimant. To discriminate between legitimate and fraudulent authentication requests, the similarity of two binary keys needs to be evaluated. Among the several metrics (such as standard error, Pearson correlation coefficient, and mutual information), our analysis exploits the fractional hamming distance (FHD) – i.e. the percentage of bits that differs between two binary strings. This is a common choice both in biometrics and in PUF characterization [6,34]. The FHD distribution between the responses to the same challenge (*like* FHD) quantifies the stability of the system, while that one between the responses to different random challenges (*unlike* FHD) is used to evaluate the correlation of the independent responses. Indeed, following the method introduced by Daugman [27,28], the number $N$ of independent bits (the entropy) of the generated keys – i.e. the number of independent degrees of freedom – can be estimated by assuming that the *unlike* FHD can be modeled with an equivalent binomial distribution $B(N, p)$ and expressed as a function of the mean value $p$ and standard distribution of the curve $\sigma$, $N =$ $\frac{p*(1-p)}{\sigma^2}$ [34,35]. We refer to intra-device FHDs when comparing responses from the same PUF or inter-device FHDs when comparing responses from different PUFs.

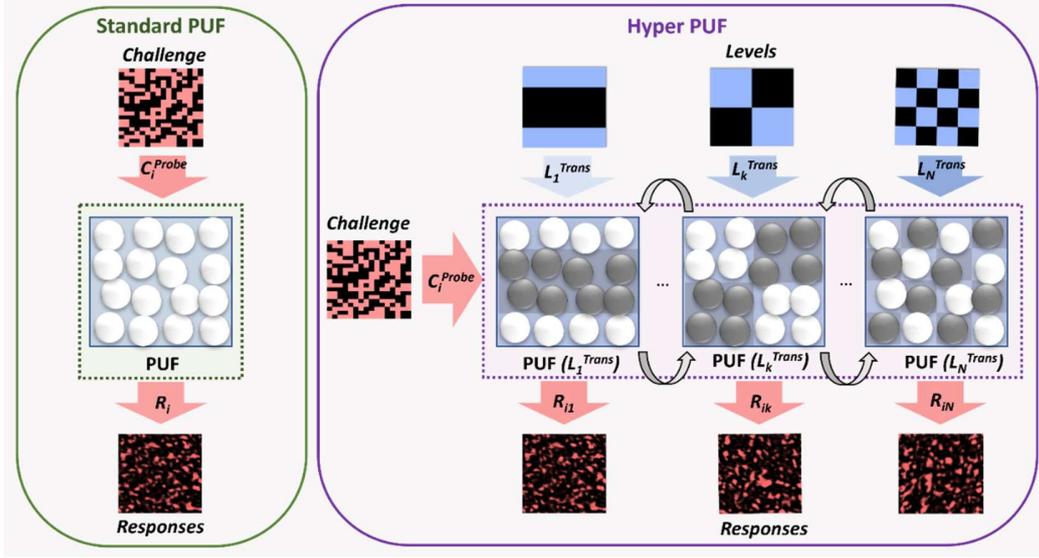

**Figure 1. Schematic representation of the interrogation process for standard and Hyper PUFs.** Working mechanism of the deterministic behavior for the challenge response pair generation for standard PUFs (left panel) and HPUFs (right panel). For the standard PUF, the challenge $C_i^{Probe}$ probes the only possible internal configuration of the hardware, producing only one response $R_i$ to a given challenge. In the HPUF, each configuration pattern reversibly transforms the PUF level into a new one, producing different responses $R_{ik}$ to a given challenge $C_i^{Probe}$.

The HPUF is a 3D disordered photonic medium that is responsive to the transformer challenge while unperturbed to the probing challenge. It consists of a polymer film where liquid crystal (LC) droplets are randomly dispersed via an emulsion process resulting in polymer dispersed and polymer stabilized liquid crystals (PD&SLC)[36] as shown in Fig. 2a. The response selectivity between the transformer and the stimulus challenges is achieved by doping the common liquid crystal 5CB with a blue absorbing dye (dispersed red 1, DR1). Blue incoherent light ($L_k^{Trans}$) transforms the internal state of the PUF by absorbing light and thereby generating a temperature driven LC phase transition, while red coherent light ($C_i^{Probe}$) probes the transformed PUFs. The LC droplets are further stabilized with cross linker molecules (Fig. 2a) that create a fixed polymeric network[37] to favor the recovery of the LC alignment in the nematic phase after the phase transition (Fig. 2b,c). Fig. 2d-f show the polarized optical microscope characterization of the nematic-isotropic-nematic phase transition of the LC within the illuminated spot of blue light. The presence of the cross linker molecules guarantees an hysteresis-free process – i.e. a reversible switching between the two LC phases[38]. The transformation between different internal configurations is deterministic, stable, and repeatable, regardless of the history of the system.

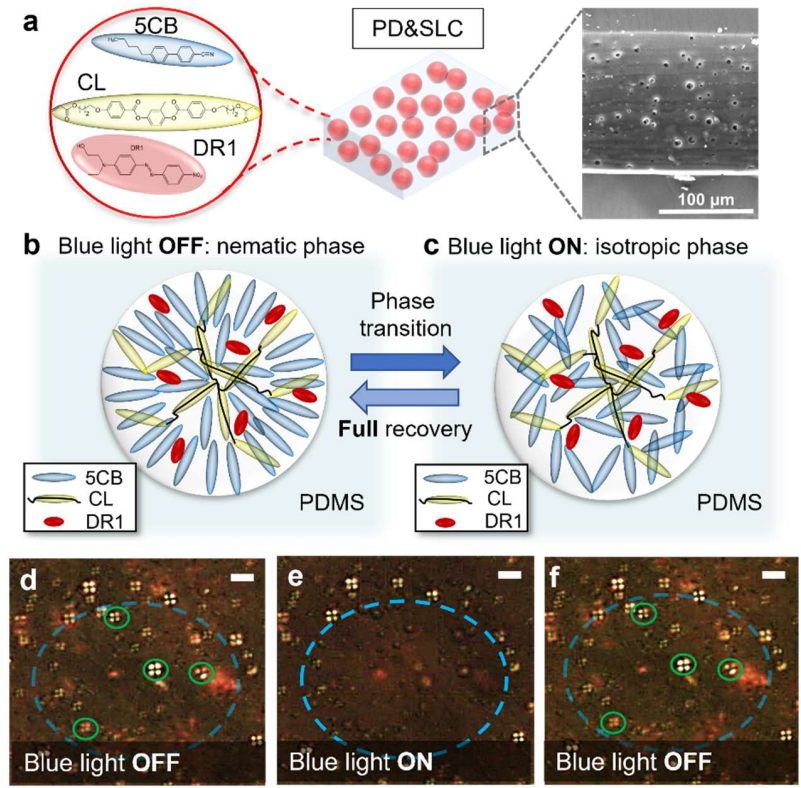

**Figure 2. Polymer dispersed and polymer stabilized liquid crystals.** a) Scheme of the polymeric film used as disordered photonic medium to realize the HPUF. The LC droplets, whose molecular composition is reported on the left, are randomly dispersed into the polymeric matrix (polydimethylsiloxane, PDMS). The polymer stabilized LC formulation is made by a mesogen (5CB), a chromophore (Dispersed Red 1, DR1), and a bi-acrylate (cross-linker) mesogen that enables a full recovery of the LC alignment. A scanning electron microscope image of the side view of the film is reported on the right. b-c) Representative scheme of the molecule arrangement inside the PD&SLC droplets for the switchable operation. b) Scheme of the LC molecular arrangement within the droplets in the nematic and c) isotropic phase. d-f) Polarized Optical Microscope images of the PD&SLC before (d), during (e) and after (f) the blue light illumination indicated by the dashed blue circle. The four dot cross pattern (d) is a signature of the LC radial alignment in each droplet[39] and it is lost under blue laser illumination. This is the indication that the stabilized LC polymer does not prevent full LC disordering to the isotropic phase. Once the blue illumination is removed (e), the system evolves in around 10 seconds to the previously aligned configuration with the same four dot feature that was present before the transformation (as highlighted by the green circles in f). The scale bars are 10 μm in length.

The experimental characterization of a HPUF is illustrated in Fig. 3a. The system is illuminated with the challenge $C_i^{Probe}$ that is generated by modulating the Gaussian wavefront of an He-Ne laser using a digital micro-mirror device (DMD) [40,41]. Light is then scattered by the HPUF, generating in the far field the response $R_{ik}$ (the speckle pattern) – a 2D image whose spatial features depend uniquely on the probe $C_i^{Probe}$ and transform challenge $L_k^{Trans}$ of the system. The response $R_{ik}$ is imaged on a CCD camera, then filtered and binarized to generate the key. The raw speckle images (the optical responses $R_{ik}$) are converted into binary keys by using a Gabor filter to remove pixel-scale noise, averaging the undesired intensity variations and extracting the independent bits[3,27]. The parameters of the Gabor filter have been tuned in order to maximize the extractable entropy from the PUF response.

Switching between the levels of the HPUF is triggered by the bright blue profile $L_k^{Trans}$ (spatially overlapped on the bright red challenge $C_i^{Probe}$), generated by a standard projector.

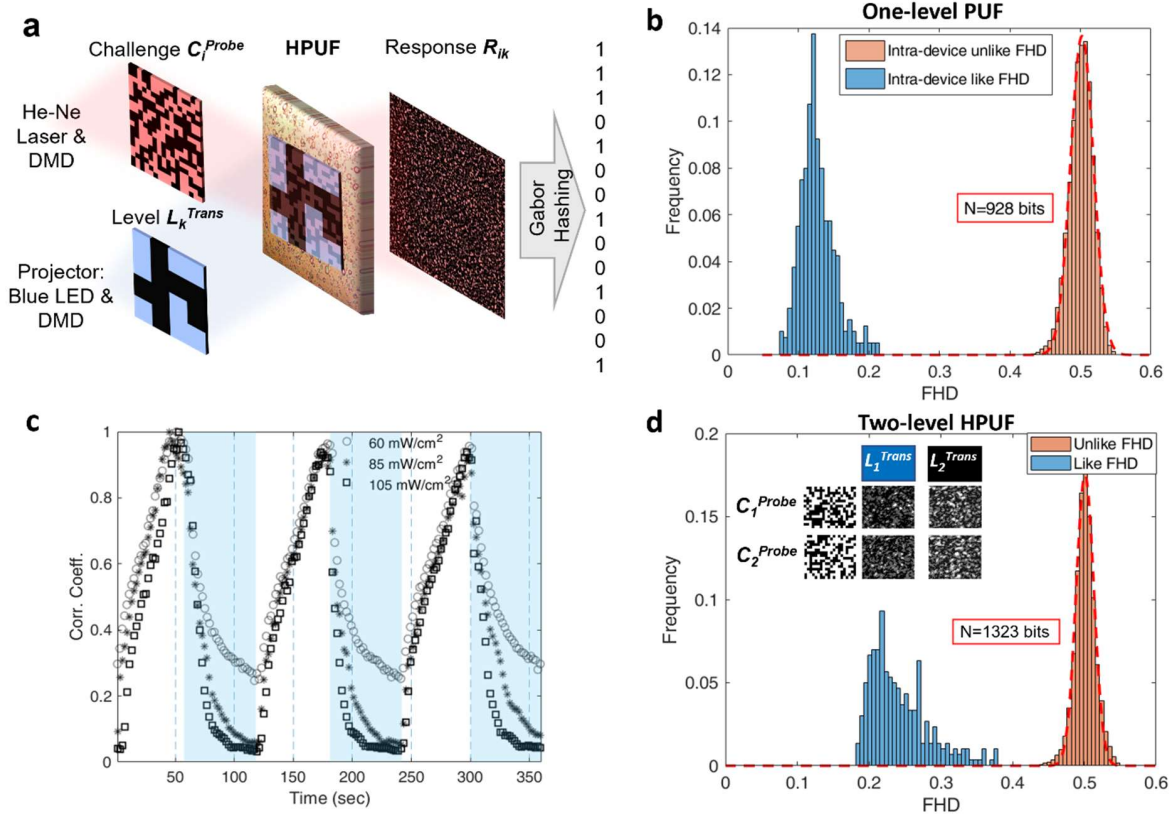

**Figure 3**. **Hyper PUF characterization: one and two-level operation.** a) Schematics of the HPUF characterization setup where a binary challenge $C_i^{Probe}$ is incident on the physical hardware. The transmitted intensity profile $R_{ik}$ is collected in the far field by a CCD camera and successively converted into a binary post-processed key. A second light beam $L_k^{Trans}$, generated by a blue LED and spatially modulated by a DMD (integrated on a projector board), is used to reversibly transform the HPUF. Among the RGB colors of the projector, the blue LED was chosen as it better matches the dye absorption peak. The two optical images are overlapped on the token. b) Characterization of one-single level of the HPUF. The mean value of the *like* FHD shows that, on the average, the 12% of the bit between two keys generated by the same challenges are different. The test has been made with 400 challenges. The *unlike* FHD (orange histogram) is the result of 79800 pairwise comparisons that results from all possible comparisons of the 400 responses to random challenges. On average, each pair of generated keys differs in the 50% of its bits and the number of the independent bits is $N_{1Level}$ =928 bits. c) Temporal evolution of the Pearson correlation coefficient between two raw responses (two speckle images) acquired every two seconds by interrogating the sample with the same challenge $C_i^{Probe}$ while switching the configuration beam between $L_1^{Trans}$ and $L_2^{Trans}$. The three curves correspond to three different values of the blue-light intensities (60, 85, 105 mW/cm$^2$), indicating that higher intensities induce a more efficient decorrelation as well as faster dynamics. We observe a full recovery of the LC alignment after every LC phase transition. d) *Like* and *unlike* FHDs of the HPUF for a two-level configuration (intensity: 85mW/cm$^2$). The number of the independent bits of the generated key is $N_{2Level}$ =1323 bits. In the inset, we report the responses for the two transformer challenges $L_1^{Trans}$ and $L_2^{Trans}$ to two different challenges, $C_1^{Probe}$ and $C_2^{Probe}$.

The characterization of the HPUF has been performed by evaluating the entropic content of the keys generated by an increasing number of levels: from one-single level PUF up to a ten-level HPUF. We firstly characterize the one-single level system by interrogating the HPUF with challenges ($C_i^{Probe}$, $L^{Trans}$=0) with i={1,…,100}. Fig. 3b shows that the *like* and *unlike* FHDs distributions are well separated, and that the authentication threshold can be safely set around 0.35. The number of the independent bits of the generated keys is estimated to be $N_{1Level}$=*928* bits.

The next step is to characterize the two-level HPUF. The first level is obtained by completely shading the blue light, while the second one is configured by illuminating the PUF with a uniform blue wavefront. The responses of the two levels interrogated with the same challenge are well decorrelated and also reproducible (see Fig. 3c). The challenge-response characterization of a two-level HPUF is done by illuminating each level with the same set of 100 random challenges (Fig. 3d). The *unlike* FHD distribution is obtained by comparing all possible pairs of responses (roughly $2*10^4$ pairwise comparisons), while the *like* FHD distribution is obtained by comparing a defined set of 150 random challenges acquired multiple times for each level. The two distributions do not overlap and the authentication threshold can be set around 0.4. We also observe a net gain in the independent bits (entropy) of the keys generated by the two-level HPUF with respect to a one-level HPUF, from $N_{1Level}$= 928 bits to $N_{2Levels}$= 1323 bits (Fig. 3b and Fig. 3d). This is the indication that two keys generated by the two-level HPUFs have a greater probability of differing in 50% of the bits (optimal situation), compared to two keys of one-level PUFs.

The natural question that arises is whether, and to which extent, a further increase in the number of levels of the HPUF increases the entropy of the generated keys. To investigate this problem, we define a set of 10 transformer challenges ($L_k^{Trans}$, with $k=\{1,..,10\}$) by choosing 10 elements of the Walsh-Hadamard binary basis – i.e. a subset of a complete 16 orthogonal set of 4x4 macropixel images. Each level is interrogated with the same set of randomly selected challenges $C_i^{Probe}$ with $i=\{1,..,100\}$. Fig. 4a shows the scheme of the domain and codomain ($R_{ik}$, with $i=\{1,..,100\}$ and $k=\{1,..,10\}$) of the HPUF. The whole codomain of responses $R_{ik}$ can be compartmentalized by randomly joining the codomains of individual levels. For each compartment, we evaluate the entropy per symbol of the keys, i.e. the entropy per bit. Fig. 4b left panel shows that the number of independent bits is of around 950 (0.14 bit/bit). This value is almost independent on the chosen level (in this case, each compartment is composed by the codomain of a single level). By populating the compartments with the codomains of more levels (up to ten), the entropy of the generated key increases up to around $N_{10Levels}$= 1750 bits that correspond to 0.24 bit/bit (Fig. 4b, left panel, red circles). The increase of the entropy per symbol evaluated by the Daugman's analysis is confirmed by modeling the extracted keys with equivalent Markov chains, generated via transition matrices whose coefficients represent the permanence and transition probabilities of the binary values of the keys. The entropy per symbol of the Markov chains is then calculated analytically[30]. The fact that the experimental data analyzed with two models show the same entropic trend suggests that the different levels behave like different cryptographic primitives coexisting in the same hardware.

To validate this idea, we fabricated ten different cryptographic primitives. We applied the same compartmentalization scheme making an analogy among each codomain of the ten different PUFs and each codomain of the ten levels of the HPUF. We observe that the entropy of the generated keys as a function of the number of PUFs, has absolute values and a dependence qualitatively similar to the HPUF (Fig. 4b, left panel, black circles). This is the confirmation that the transformer challenges $L_k^{Trans}$ induce different microscopic configurations in the same region of the sample, mimicking different PUFs. It is important to notice that the entropy increase does not depend on the number of responses of the PUFs but only on the number of levels of the HPUF. Increasing the number of responses but considering a single configuration, the entropy per symbol of the PUF remains constant.

The increase in entropy per symbol implies a greater unpredictability of the bit sequence. By analyzing the properties of the equivalent Markov chains, we observe that the increase of the number of levels leads to permanence and transition ($\alpha$ and $1-\alpha$ respectively) probabilities of the Markov transition matrix, that tend towards a situation of equiprobability ($\alpha = 0.5$). This implies the reduction of the correlation length in the bit sequence (Fig. 4c-d). Indeed, the correlation length of the bit sequence gets shorter and shorter when increasing the number of levels (Fig. 4c), until it reaches an asymptotic value and the entropy per symbol saturates. The physical origin of the increase in the entropy per symbol is due to an increases of the microscopic configurations of the system probed by the challenge $C_i^{Probe}$, that translates in an increase of the variety of speckle patterns that form the codomain of the HPUF. The growing rate of the entropy is reduced up to a saturation level when the compartment is populated by roughly 8-10 levels. For a given size of the challenge $C^{Probe}$ and $L^{Trans}$, the entropy per symbol saturates when light probes all the possible accessible configurations of the system.

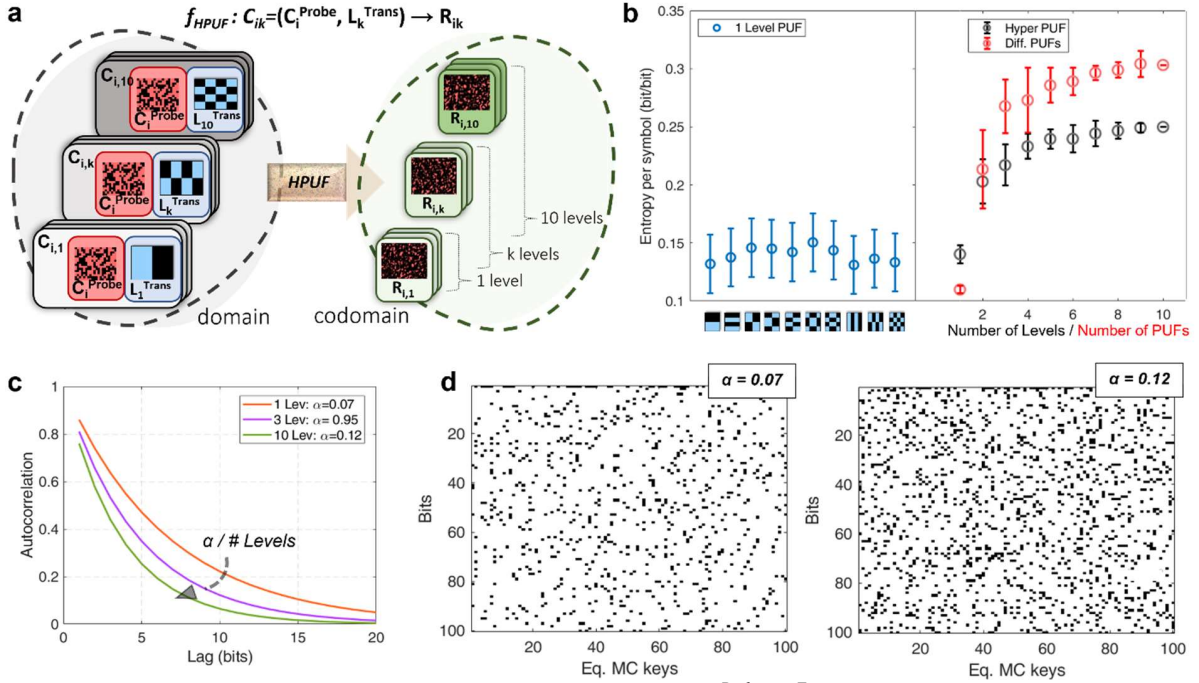

**Figure 4. Hyper PUF characterization.** a) Scheme of the domain ($C_i^{Probe}$, $L_k^{Trans}$) and the codomain $R_{ik}$ of the HPUF. The whole codomain can be compartmentalized by joining the responses of different levels. b) Entropy per symbol for different compartments of the HPUF codomain. The left panel (blue circles) shows the entropy per symbol for each single level (the horizontal labels show the Hadamard basis configuration patterns). The right panel shows the increase of the entropy per symbol by randomly joining the codomains of individual levels (black circles). Red circles refer to the same analysis performed by populating the compartment by joining the responses of different PUFs. The blue error bars refer to 10 different characterizations of each single level. The black and red error bars refer to the standard deviation calculated over ten different random selections of the PUF or levels of HPUF, respectively. c) Autocorrelation of the bit sequences generated by equivalent Markov chain. The correlation length decreases as the number of levels increases, because the permanence probability of the equivalent key increases from $\alpha = 0.07$ to $\alpha = 0.12$. d) Representation of the binary keys generated by the equivalent Markov chains for one level ($\alpha = 0.07$) and ten levels ($\alpha = 0.12$).

## Conclusions

We developed new optical cryptographic primitives, named Hyper PUFs or HPUFs, that allow multi-level operation thanks to fully reversible switching of their optical properties. The all-optical HPUF of this paper is realized with polymer dispersed and stabilized liquid crystals, and the transformation of the levels is enabled by a light pattern that can selectively and locally drive the reversible phase transition of the embedded liquid crystals. The entropy of the HPUF's keys has been studied using different methods, both confirming its increase with the number of joint levels. These results show that the HPUF is equivalent to combining several physical unclonable functions into a single hardware. The overall entropy per bit is affected by the unavoidable presence of spatial correlations between the microscopic configurations and reaches a saturation levels when the challenge light probes all the microscopic configurations of the system. We believe that the concept described in this paper allows for the development of a new generation of all-optical security devices. Amongst the advantages is the unique possibility to create multi-level PUFs integrated into one and the same material, thus enabling a practical implementation of quantum secure authentication of data[10]. This not only opens up to novel quantum protocols via strong optical PUF but significantly increases their security level and also allows to have multi-user key generators and hence multiple clients on one device. [22,33]

## Acknowledgements

The research leading to these results has received funding from Ente Cassa di Risparmio di Firenze (2018/1047), AFOSR/RTA2 (A.2.e. Information Assurance and Cybersecurity) project "Highly Secure Nonlinear Optical PUFs" (Award No. FA9550-21-1-0039) and Fondo premiale FOE to the project "Volume photography: measuring three dimensional light distributions without opening the box" (E17G17000300001). The authors thank Prof. Hui Cao, Yaniv Elisier, Giuseppe EmanueleLio, Micheal Lachner, Nils Wisiol and Adomas Baliuka for stimulating discussion.